\documentclass[fleqn,usenatbib]{mnras}

\usepackage{newtxtext,newtxmath}

\usepackage[T1]{fontenc}
\usepackage{ae,aecompl}
\usepackage{xcolor}

\usepackage{graphicx}	
\usepackage{amsmath}	
\usepackage{amssymb}	
\usepackage{ulem}       

\usepackage{array}      
\usepackage{makecell}   

\title[Reverberation lag in Mrk~335]{Multi-timescale reverberation mapping of Mrk~335}
\author[G. Mastroserio et al.]{
Guglielmo Mastroserio,$^{1,2}$\thanks{E-mail: gullik@caltech.edu }
Adam Ingram,$^{3}$
\& Michiel van der Klis$^{2}$
\\
$^{1}$Cahill Center for Astronomy and Astrophysics, California Institute of Technology, 1200 California Boulevard, Pasadena, CA 91125, USA \\
$^{2}$Astronomical Institute Anton Pannekoek, University of Amsterdam, Science Park 904, NL-1098 XH Amsterdam, Netherlands\\
$ ^{3} $Department of Physics, Astrophysics, University of Oxford, Denys Wilkinson Building, Keble Road, Oxford OX1 3RH, UK
}

\date{Accepted XXX. Received YYY; in original form ZZZ}

\pubyear{2019}

\begin{document}
\label{firstpage}
\pagerange{\pageref{firstpage}--\pageref{lastpage}}
\maketitle

\begin{abstract}
Time lags due to X-ray reverberation have been detected in several Seyfert galaxies. The different travel 
time between reflected and directly observed rays
naturally causes this type of lag, which depends directly on the light-crossing timescale of the system and hence scales with
the mass of the central black hole.
Featureless `hard lags' not associated with reverberation, and often interpreted as propagating mass accretion rate fluctuations, dominate the longer timescale variability. Here we fit our \textsc{reltrans} model simultaneously to the time-averaged energy spectrum and the lag-energy spectra of the Seyfert galaxy Mrk~335 over two timescales (Fourier frequency ranges). We model the hard lags as fluctuations in the slope and strength of the illuminating spectrum, and self-consistently account for the effects that these fluctuations have on the reverberation lags.
The resulting mass estimate is  $1.1^{+2.0}_{-0.7} \times 10^6~M_\odot$, which is significantly lower than the mass measured with the optical reverberation mapping technique ($14$ - $26$ million $M_{\odot}$). 
When we add the correlated variability amplitudes to the time lags by fitting 
the full complex cross-spectra, the model is unable to describe the characteristic reverberation Fe K$\alpha$ line and cannot constrain the black hole mass. 
This may be due to the assumption that the direct radiation is emitted by a point-like source.

\end{abstract}




\section{Introduction}

Active Galactic Nuclei (AGN) are thought to be  powered by the accretion of matter onto 
supermassive black holes ($10^6 - 10^9 M_{\odot}$). 
The gas forms an accretion disc and emits a multi-temperature blackbody spectrum which peaks in 
the UV band (\citealt{Shields1978,Malkan1983}). Some of these photons act as seed photons for inverse Compton up-scattering in a hot electron `corona' (\citealt{Eardley1975, Thorne1975}). 
The resulting \textit{direct} Comptonised emission is often simply modelled as a power-law with a high energy cut-off, where the power-law index and 
the cut-off energy are related to the optical depth and the electron temperature, respectively. 

Part of this radiation illuminates the disc and is re-emitted, producing the \textit{reflection} component in the spectrum 
 (e.g., \citealt{Lightman1988}).  
The shape of the overall time-averaged energy spectrum depends on the properties of both the corona, such as optical depth and temperature, and the accretion disc, including both its physical characteristics  
(e.g., ionisation and iron abundance) and its geometry (e.g., inclination and radius of the inner edge). 
One of the most prominent features in the reflection spectrum is an iron $K\alpha$ line emitted at $6.4$ keV and broadened by the effects of orbital motion in the disc and relativistic redshift (\citealt{Fabian1989}).  The black hole drags space-time around it further modifying the emission line profile (\citealt{Fabian2000}), 
making it possible to measure the spin of the black hole (see  \citealt{Reynolds2014} for a review). 

Although this approach of time-averaged spectral fitting 
led to very interesting results, there are several 
 aspects that have remained unclear, in particular, the exact geometry of the system. 
For example, the extent of the corona is under debate. It might be either a compact cloud of gas that could reasonably (compared to the accretion disc dimensions) be 
modelled as a point source (\citealt{Haardt1991}), or 
a more extended structure, with various geometries being considered for it (e.g. \citealt{Eardley1975,Haardt1993}). 
One of the reasons for this ambiguity is that in time averaged spectral fitting, degeneracies among the model parameters can often not be avoided. 
In order to help break such degeneracies, it is possible to additionally study the time variability of the spectrum. 
One effective way to study the spectral variability is to measure the time lags observed  
in both supermassive and stellar mass black holes between the variability in different photon energy bands.
Two types of lag are observed in AGN, the `hard lag' that is intrinsic to the direct emission (intrinsic lags)
and the `soft lag' that is due to the differences in light crossing time from corona to observer between the direct and the reflected emission (the reverberation lags). 
The intrinsic lags are thought to be generated by mass accretion rate fluctuations propagating
through the accretion disc towards the black hole (e.g. \citealt{Lyubarskii1997,Arevalo2006}; \linebreak \citealt{Ingram2013}). 
The timescales of the fluctuations depend on the 
viscosity of the gas, thus long timescales arise in the outer part of the accretion flow and short timescales closer to the black hole. The fluctuations reach the corona causing variations in the electron temperature.  
At each radius the variability is due to the product of local fluctuations 
and the generally slower fluctuations that propagated in from larger radii. 
Conversely, the reverberation lags are due to the different path 
lengths of the variability signals to reach the observer. 
Hence, the variability in the energy band dominated by the reflected emission 
lags behind the variability of the direct, inverse Compton emission. 

The hard lags dominate at long timescales, and hence in 
many sources the lags as a function of Fourier frequency computed between $\simeq 0.3-1.0$ keV (soft energy band)
and $\simeq 2.0-4.0$ (hard energy band) are positive (i.e., hard flux lags soft flux) at Fourier frequencies $\nu \leq 300 M_{\odot}/M$ Hz. The lag spectrum as a function of photon energy calculated in this frequency range is featureless and lag depends approximately linearly on $\log E$ (e.g. \citealt{Papadakis2001,McHardy2004}).
At higher frequencies the hard intrinsic lags are smaller, and soft (negative) lags have been detected between the same two energy bands (e.g. \citealt{Fabian2009}). 
The lag energy spectrum in this higher frequency range is indicative of reverberation, exhibiting a negative slope for $E \lesssim 4$ keV (hence the soft lag between the two broad energy bands), with a broad iron line feature in the $6-7$ keV range (e.g. \citealt{Kara2016}) and a Compton hump in the $10-80$ keV range (\citealt{Zoghbi2014}).
Detecting this Fe line reverberation feature in the lag energy spectrum 
is easier in AGN than in stellar mass black holes
because they are much bigger systems and 
so the timescales involved are consequently longer.
However, because of their distance, AGN are usually fainter than stellar mass 
black holes, so they are often characterized by a worse signal to noise. 
For that reason, we need to use as wide as possible a frequency range to calculate the lag energy spectrum, and analyse it together 
with the time-averaged energy spectrum in order to constrain the system parameters. 

Looking only at the time lags means considering only a limited part of the information in the data, as the
 correlated variability amplitude as a function of energy and frequency is ignored in this approach. 
Lags and amplitudes have been considered together in only a few cases (e.g. \citealt{Uttley2011, Kara2013a, Rapisarda2016}), however, joint modelling of both has not yet been performed for AGN.
So, progress could be made by jointly modelling the time-averaged and variability (time lag and correlated amplitude) spectra taking into account both the effects of  mass accretion rate propagating 
fluctuations and reverberation lags. The best way to consider time lags and correlated amplitudes together is by explicitly modeling the real and imaginary parts of the complex cross spectra (\citealt{Klis1987, Rapisarda2016}; \citealt{Ingram2016,Mastroserio2018}),
and that is the approach taken in this paper.

We have developed the fully relativistic reverberation mapping model 
\textsc{reltrans} (\citealt{Ingram2019}) that, using a transfer function formalism (e.g. \citealt{Campana1995, Reynolds1999, Cackett2014}), computes the complex cross-spectrum 
for a range of Fourier frequencies and the time-averaged energy spectrum in a
prescription based on 
a `lamppost' coronal geometry (e.g. \citealt{Matt1991}),
assuming isotropic emission and a flat thin accretion disc. 
In the configuration of \textsc{reltrans} that we use, the disc radial ionisation profile is self-consistently calculated from the radial density profile
corresponding to a \citet{Shakura1973} zone~A accretion disc. 
However, the rest-frame reflection spectrum is computed using the model xillver 
(\citealt{Garcia2013}), which employs a fixed electron number density of 
$10^{15}~{\rm cm}^{-3}$. Future versions of reltrans will include the variable electron 
density version of \textsc{xillver}, \textsc{xillverD} (\citealt{Garcia2016}), which will enable us to include self-consistently a 
radial \textit{density} profile as well as a radial ionization profile.
The \textsc{reltrans} model computes the reverberation lags and accounts for the 
different light crossing times of the photons using proper relativistic ray-tracing both between corona and each point on the disc and from each point on the disc to the observer. 
We use here a further improved version of \textsc{reltrans} which
accounts for the hard lags through the pivoting of the continuum spectrum
by allowing fluctuations in the index and normalisation of the power law  (\citealt{Mastroserio2019}).
This serves as a proxy for the cooling and heating of the corona by the fluctuations in the rate of seed photons coming from 
the disc and the mass accretion rate fluctuations propagating into the corona.  
Each patch of the disc sees a different hardness of the incident emission coming from the corona due to the effects of orbital motion and redshift which are properly evaluated taking into account 
the different paths of the photons (\citealt{Mastroserio2018}). 
This causes non-linear effects in the fluctuations of the reflected emission (i.e., not just the strength but also the shape of the reflection spectrum fluctuates) which are taken into account in the model. 
\textsc{reltrans} has been successfully tested on Cygnus~X-1 (\citealt{Mastroserio2019})
and previously used for a proof of principle of the method (fitting only the reverberation lags) in Mrk~335 (\citealt{Ingram2019}). 

In this paper we present a full exploration of the model with an XMM-Newton observations of Mrk~335. 
The Seyfert 1 galaxy Mrk~335 has been extensively studied both in terms of time-averaged spectrum 
(e.g. \citealt{Wilkins2015, Keek2016}) and lag spectrum, which shows both hard and soft lags
(e.g., \citealt{Kara2013c, DeMarco2013b, Chainakun2016a}). 
The source has been observed at different flux levels (defining 'flux epochs') (\citealt{Wilkins2013}) with XMM-Newton from 2006 to 2009. 
Reverberation lags are detected only in the highest-flux epoch (\citealt{Kara2013c}) although the reflection component in the spectrum is weaker then than in the low flux epochs (\citealt{Keek2016}). 
We therefore select the high flux epoch and use \textsc{reltrans} to  simultaneously fit 
the time-averaged energy spectrum and the complex cross-spectra as a function of energy 
in multiple Fourier frequency ranges.

\section{Data reduction}

We analyse a $133$ ks observation of Mrk~335 performed with \textit{XMM-Newton} in 2006 (obs ID  0306870101).
During the observation the source was in the high-flux epoch (\citealt{Wilkins2015}) 
with a $0.5-10$ keV flux of $4.08 \times 10^{-11}$ erg $\mathrm{cm}^{-2}$ $\mathrm{s}^{-1}$ \citep{Grupe2007}. 
We follow the data reduction procedure described in \cite{Ingram2019}, 
considering only the EPIC-pn data and discarding the EPIC-MOS data (following \citealt{Kara2013c}). We use the Science Analysis System (SAS) v11.0.0 to extract the signal from a circular region with $35$ arcsec radius centred on the maximum of the source emission. 
We apply the filters PATTERN $\leq 4$ and FLAG == $0$ and discard background 
flares at the beginning and at the end of the observation 
(considering only times 252709714 to 252829414 seconds telescope time). 
We extract the time-averaged source and background energy spectrum and 
also extract light curves with $10$ second binning from $12$ different 
energy bands with the same energy resolution as used by \citet{Kara2013c}. 
The SAS task \textsc{epiclccorr} performs various corrections including 
subtracting the background signal, which is extracted from a circular 
region of $35$ arcsec radius at some distance from the source.

We compute the cross-spectrum between the light curve for each energy band and the reference light curve which is the sum of all the light curves except the subject one (\citealt{Uttley2014}), and average it into broad frequency ranges. 
We consider the full length of the light curves in order to probe the same frequency ranges as \citet{Kara2013c} and \citet{DeMarco2013b}.
Before Fourier transforming
the light curves, we fill in small gaps (less than $100$ seconds) by
randomly extracting the missing count rates from a Poisson distribution centered on the value interpolated between the previous and the next bin of the gap.
We divide the cross-spectrum into two frequency ranges: $ 0.02 - 0.2 $ mHz, $0.2 - 0.7 $ mHz. 
These two frequency ranges are roughly in the regimes dominated by  
the hard and soft lags, respectively,  
as found by \citet{DeMarco2013b}.
We find that the intrinsic coherence (\citealt{Vaughan1997}) is consistent with unity in these two frequency ranges but drops off significantly at higher frequencies, as is seen in other AGN (\citealt{Zoghbi2010, Uttley2014}). Since the model assumes unity coherence, we ignore the higher frequencies that display low coherence.
Since the Fourier frequency resolution of the cross-spectrum is $ 8.35 \times 10^{-6}$ Hz, the first frequency range contains $21$ Fourier frequencies, and the second $60$. The cross-spectral amplitudes averaged over these two frequency ranges are therefore sufficiently close to Gaussian-distributed 
that we can use the $\chi^2$ statistic for the purposes of fitting models (\citealt{Nowak1999}).

\citet{Epitropakis2016a} pointed out that considering the full length of the light curve and averaging the cross-spectrum into broad frequency bins (smoothing) can introduce a bias to the time-lag estimates. Specifically, the time lag estimated for a given frequency bin is not equal to the expectation value of the true time lag at the central frequency of the considered frequency bin.
This occurs because the amplitude and phase of the cross-spectrum can depend quite steeply on frequency, and therefore the cross-spectrum can change quite substantially and non-linearly within a broad frequency bin.
The improved version of \textsc{reltrans} that we present in this paper (see Section~\ref{sec:reltrans}) models this bias by first calculating the predicted frequency dependent cross-spectrum and then averaging over the frequency range corresponding to the observational data. 
This can be understood as the model including exactly the same biases as the data do, leading to the best fitting model parameters themselves being unbiased\footnote{We note that, although this fit procedure is unbiased, it does ignore diagnostic information. As an extreme example, if the phase turns by $360^\circ$ in  a frequency bin, the cross-spectrum of data and model can 
be zero for that bin. In such an example, the fit would not be biased, but it \textit{would} be degenerate.}. 
The old version of the model (\citealt{Ingram2019}\footnote{The old version of the code is publicly available at \url{www.adingram.bitbucket.io/reltrans.html}}) onlyapproximately followed the described procedure. These approximations are good for the low frequencies (in terms of $c/R_g$) probed by fits to X-ray binary data (e.g. \citealt{Mastroserio2019}), but break down for the higher frequencies (again in terms of $c/R_g$) probed by fits to AGN data. We therefore first address the inaccuracies in the old model (Section~\ref{sec:reltrans}) before fitting to the Mrk 335 data (Section~\ref{sec:fits}).

\section{\textsc{reltrans} model improvements}
\label{sec:reltrans}
In this Section, we describe the new version of the \textsc{reltrans} model used in this paper. A future paper will detail further improvements in addition to those described here, and will be accompanied by the public release of a new model version (version 2). The initial release of the model was described in \citet{Ingram2019}. 
Subsequently,  \citet{Mastroserio2019} included the non-linear effect of variations in the power-law index of the illuminating spectrum, which can reproduce the observed hard lags. Here, we build on the \citet{Mastroserio2019} model by relaxing some assumptions that caused inaccuracies in the high frequency ranges (in terms of $c/R_g$) probed by AGN. 
Averaging the transfer functions over the frequency range (the method used in the old version of the model) is correct only when the transfer function (modulus and phase) and the variability amplitude of the continuum variations do not depend strongly on frequency. This is not the case
for the higher frequencies at which
the reverberation lags dominate.
Thus the improved model calculates the frequency dependent cross-spectrum before then averaging over the specified frequency range.
We present below the mathematical details of our improvements to the model, which include a slight change in the model parameters specifying the spectral pivoting that makes them easier to interpret physically.

\subsection{Formalism}

We start by representing the time-dependent spectrum directly observed from the corona as $D(E,t) = A(t) \ell g_{\rm soz}^{\Gamma(t)} E^{1-\Gamma(t)} {\rm e}^{-E/E_{\rm cut,obs}}$, where $\ell$ and $E_{\rm cut, obs}$ are respectively the lensing factor and observed high energy cut-off as defined in \citet{Ingram2019} and $g_{soz}\equiv g_{\rm so}/(1+z)$, where $g_{\rm so}$ is the blueshift experienced by a photon traveling from the corona to an observer at infinity in an asymptotically flat and static 
spacetime and $z$ is the cosmological redshift experienced by the photon. The normalisation $A(t)$ varies around a mean value $A_0$ and the power-law index $\Gamma(t)$ varies (with small variability amplitude) around a mean value $\Gamma_0$. The total spectrum is the sum of  the directly observed emission and the reflected emission, which we calculate using the transfer function formalism extensively described in \citet{Ingram2019} and \citet{Mastroserio2019}.

The Fourier transform (FT) of the full spectrum is
\begin{align}
    S(E,\nu) = & {\rm e}^{-\tau(E)} \bigg\{ A(\nu) \left[ D(E) + W_0(E,\nu) \right] \nonumber \\
    &+ A_0 \Gamma(\nu) \left[ \ln(g_{\rm soz}/E) D(E) + W_1(E,\nu) + W_2(E,\nu) \right] \bigg\}.
    \label{eqn:senu00}
\end{align}
This expression is very similar to Equation 8 in \citet{Mastroserio2019}, except here we explicitly include interstellar absorption with the ${\rm e}^{-\tau(E)}$ factor (this was left as implicit in \citealt{Mastroserio2019}), and \citet{Mastroserio2019} defined a new term $B(\nu)=-A_0 \Gamma(\nu)$, which we now refrain from doing in order to make the model parameters easier to interpret. Here, $D(E)\equiv \ell g_{\rm soz}^{\Gamma_0} E^{1-\Gamma_0} {\rm e}^{-E/E_{\rm cut,obs}}$, and $W_0(E,\nu)$, $W_1(E,\nu)$ and $W_2(E,\nu)$ are transfer functions that are written out in full in Appendix \ref{sec:transfer}. $W_0$ represents the response of the reflection spectrum to fluctuations in the normalisation of the illuminating spectrum and $W_1$ and $W_2$ represent the response to fluctuations in the power-law index ($W_1$ accounts for changes to the emissivity profile and $W_2$ accounts for changes to the restframe reflection spectrum).

Defining $\gamma(\nu) {\rm e}^{i\Delta_{\rm BA}(\nu)} \equiv A_0 \Gamma(\nu)/A(\nu)$, Equation (\ref{eqn:senu00}) becomes
\begin{align}
    S(E,\nu) = & A(\nu) {\rm e}^{-\tau(E)} \bigg\{ D(E) + W_0(E,\nu)
    + \gamma(\nu) {\rm e}^{i\Delta_{\rm BA}(\nu)} \nonumber\\
    &\left[ \ln(g_{\rm soz}/E) D(E) + W_1(E,\nu) +  W_2(E,\nu) \right] \bigg\}.
    \label{eqn:senu}    
\end{align}
Here $\gamma(\nu) = A_0 |\Gamma(\nu)|/|A(\nu)   |$ is the amplitude of fluctuations in $\Gamma$ divided by the fractional variability amplitude of fluctuations in $A$, 
and $\Delta_{\rm BA}(\nu)$ is the phase lag between fluctuations in $\Gamma$ and those in $A$ (positive means $\Gamma$ lags $A$). 
Since it is easier to interpret, we adopt $\Delta_{BA}(\nu)$ in place of $\phi_B(\nu)$, which we used in \citet{Mastroserio2018} and \citet{Mastroserio2019}.

For the case of the observational data, we take the Fourier transform (FT) of the count rate in the subject band and then cross with the FT of the reference band count rate. The model has to follow the same procedure. In previous versions of the model, we averaged the transfer functions over the
frequency range specified as model parameters $\alpha(\nu)=|A(\nu)|$, $\phi_A(\nu)={\rm arg}[A(\nu)]$ and $\phi_B(\nu)$ averaged over the same frequency range. 
This procedure is inaccurate if any of these three quantities, or any of the transfer functions, change significantly across the frequency range. 
We therefore now first calculate the frequency dependent cross-spectrum and then average over the input frequency range.

The cross-spectrum is $G(E,\nu)=S(E,\nu)F_r^*(\nu)$, where $F_r(\nu)$ is the FT of the reference band count rate time series. The reference band count rate is the sum of the count rates from each of the energy channels $I_{\rm min}$ to $I_{\rm max}$ that make up the reference band.
To calculate the reference band FT, we must therefore fold $S(E,\nu)$ around the instrument response to get $S(I,\nu)$ and then sum over the reference band channels. The cross-spectrum is therefore
\begin{equation}
    G(E,\nu) = |A(\nu)|^2 S_{\rm raw}(E,\nu)F_{\rm r,raw}^*(\nu) = |A(\nu)|^2 G_{\rm raw}(E,\nu),
\end{equation}
where $S_{\rm raw}(E,\nu) \equiv S(E,\nu)/A(\nu)$ and $F_{\rm r,raw}(\nu) \equiv F_{\rm r}(\nu)/A(\nu)$.
Note that $G(E,\nu)$ can be very simply folded around the instrument response to get $G(I,\nu)$.

Once we have calculated the cross-spectrum for the data, we average over a frequency range $\nu_{\rm lo}$ to $\nu_{\rm hi}$ centered at $\nu_c = (\nu_{\rm hi} + \nu_{\rm lo})/2$. Again the model has to do the same to get
\begin{equation}
    \langle G(E,\nu_c) \rangle = \frac{ \int_{\nu_{\rm lo}}^{\nu_{\rm hi}} G(E,\nu) d\nu } { \nu_{\rm hi} - \nu_{\rm lo}  } = \frac{ \int_{\nu_{\rm lo}}^{\nu_{\rm hi}} |A(\nu)|^2 G_{\rm raw}(E,\nu) d\nu } { \nu_{\rm hi} - \nu_{\rm lo}  }.
\end{equation}
We therefore need to assume a form for $|A(\nu)|^2$. We represent this as a power-law function of frequency, $|A(\nu)|^2 = \alpha(\nu_c) (\nu/\nu_{\rm c})^{-k}$, giving
\begin{equation}
    \langle G(E,\nu_c) \rangle  = \alpha(\nu_{\rm c}) \frac{ \int_{\nu_{\rm lo}}^{\nu_{\rm hi}} (\nu/\nu_c)^{-k} G_{\rm raw}(E,\nu) d\nu } { \nu_{\rm hi} - \nu_{\rm lo}  }.
\end{equation}
The full band power spectrum is $P(\nu) \sim |A(\nu)|^2$, and observationally tends to consist roughly of a twice broken power-law: $k \approx 0$ at very low frequencies ($\nu \lesssim 0.1$ Hz in X-ray binaries), $k \approx 1$ at intermediate frequencies and $k \approx 2$ at high frequencies ($\nu \gtrsim 5$ Hz in X-ray binaries). The model is insensitive to $k$ in the low and intermediate frequency ranges for which $k\approx 0$ and $k\approx 1$. This is because for such values of $\nu_c$, $1/\nu_c$ is much longer than the light-crossing timescale of the disc, and therefore the transfer functions do not change much from $\nu=\nu_{\rm lo}$ to $\nu=\nu_{\rm hi}$. The model \textit{is} sensitive to $k$ for the highest frequencies, because here the cross-spectrum (modulus and phase) can vary steeply with $\nu$. We therefore adopt $k=2$.

The output model for the cross-spectrum is therefore $\langle G(E,\nu_c) \rangle$ (real and imaginary parts). The model lag-energy spectrum is given by $t_{\rm lag}(I,\nu_c) = {\rm arg}[\langle G(I,\nu_c) \rangle] / (2\pi \nu_c)$, which exactly mirrors our procedure to constrain the lag-energy spectrum from the observed cross-spectrum. The model parameters for the hard lags are $\gamma(\nu_c)$ and $\Delta_{BA}(\nu_c)$. Note that the lag-energy spectrum model is \textit{not} sensitive to the parameter $\alpha(\nu_c)$, but the model for the real and imaginary parts of the cross-spectrum is because $\alpha(\nu_c)$ affects the modulus of the cross-spectrum.

\section{Spectral timing analysis}
\label{sec:fits}

\subsection{Time-averaged energy spectrum}
\label{sec:DC} 
\begin{figure}
	\includegraphics[width=\columnwidth]{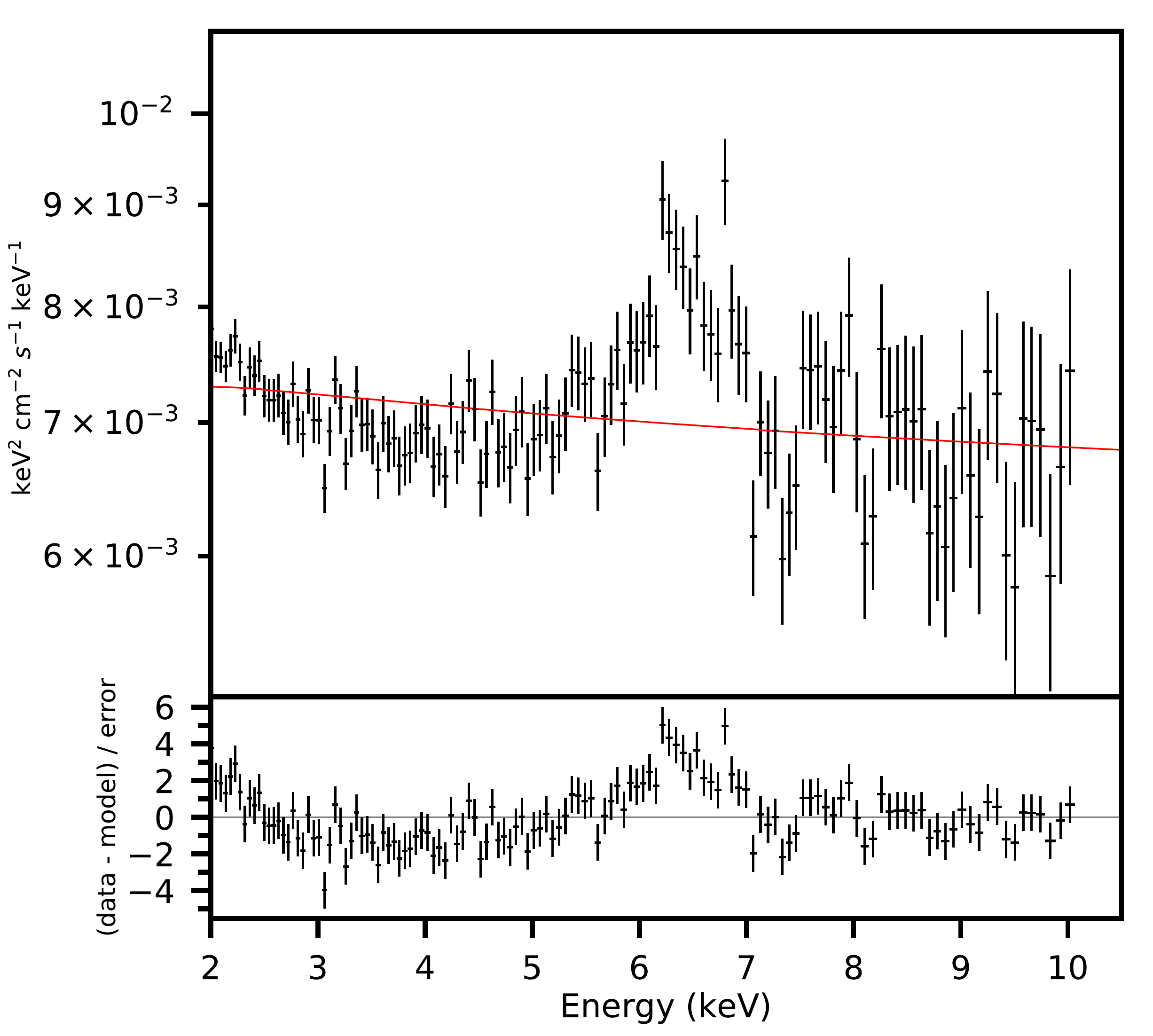}
    \caption{Upper panel Time-averaged energy spectrum unfolded with an absorbed power-law model. Lower panel: residuals of this model. 
    Clear residuals around the iron line energy range can be seen.}
    \label{fig:del_pw}
\end{figure}

Fitting the $2-10$ keV time-averaged spectrum with an absorbed power-law model 
reveals evident residuals in the iron $K\alpha$ line energy range (see Fig.~\ref{fig:del_pw}), 
consistent with previous analyses \citep{Keek2016,Wilkins2015}. 
We also see the emission line at $7.01$ keV in the rest frame of the host galaxy (the cosmological redshift to Mrk 335 is $z=0.025785$; \citealt{Huchra1999}) that was reported by \citet{Keek2016}. Throughout this paper, we model this line in the time-averaged spectrum with a narrow Gaussian function with fixed width ($10^{-3}$) and free centroid (allowed
to vary between $6.9$ and $7.1$). Including a \textsc{xillver} \citep{Garcia2013} component in the model to account for distant reflection leads to a fit with $\chi^2/{\rm d.o.f.}$ of $167.6/127$ (Model [A] in Table~\ref{tab:parameters}). 
This poor fit implies that a relativistic reflection component is also required, for which we use the model \textsc{reltrans} \citep{Ingram2019}. 
The full expression for our model (Model [B]) is
\begin{equation}
    \textsc{tbabs} \times ( \textsc{xillver} + \textsc{reltrans} + \textsc{zagauss} ),
\end{equation}
where the direct continuum emission is included in the \textsc{reltrans} model as an exponentially cut-off power-law. 
We fix the hydrogen column density to  $n_{\rm H} = 3.6 \times 10^{20} {\rm cm}^{-2}$ following \cite{Kalberla2005} and assume the relative elemental abundances of \cite{Wilms2000}. 
We fix the ionization parameter of the distant reflector to $\log\xi=0$. In contrast, we model the radial profile of the disc ionization parameter using a self-consistent calculation of the irradiating flux in a lamppost geometry and assuming the radial density profile that corresponds to `zone A' in the \cite{Shakura1973} disc model
\footnote{We note that relativistic corrections are ignored in the Shakura \& Sunyaev model. In our model the radial derivative is the relevant quantity because the assumed density profile only influences the radial ionization \textit{gradient}, not the absolute value, which is set by the $\log\xi_{\rm max}$ model parameter. However, the relativistic corrections can change the density gradient and we plan to include these corrections in a future version of the model.}. 
We use 10 radial zones to calculate the ionisation profile: 
one zone between $r_{\rm in}$ and $(11/9)^2 r_{\rm in}$ and the rest logarithmically separated. $r=(11/9)^2 r_{\rm in}$ is approximately the radius whereby the ionisation parameter peaks\footnote{It is exact if the illuminating flux scales as $r^{-3}$. We use this radius instead of numerically calculating where the ionization parameter peaks to save computational expense.},
and the value of $\log\xi$ at this radius is a model parameter that we leave free in the fit.
We choose to only model the $2.0-10$ keV energy range of the spectrum because the $0.3-2.0$ keV energy range can only be adequately described by including a complicated 3 layer warm absorber model \citep{Longinotti2013}, which is beyond the scope of this paper.

\begin{figure}
	\includegraphics[width=\columnwidth]{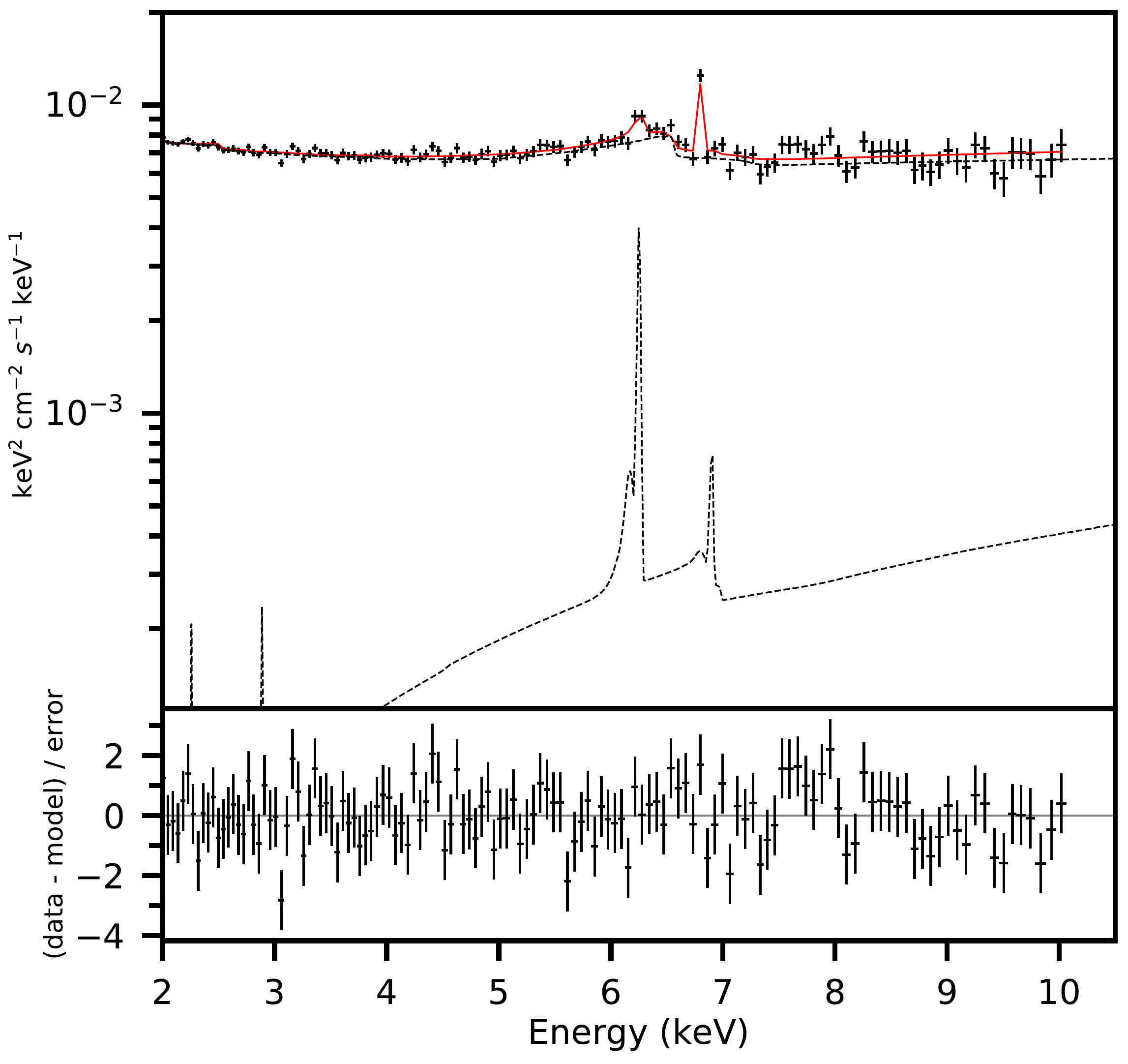}
    \caption{Upper panel: Time-averaged energy spectrum unfolded with the best fitting model accounting for direct emission and relativistic component (\textsc{reltrans}) and the distant reflector (\textsc{xillver}). The narrow emission line at $7.01$ keV in the source restframe is modelled with a Gaussian component. Lower panel: Residuals of the best fit model. }
    \label{fig:fit_he_rel_xil_zg_ion1_errors}
\end{figure}

Fig~\ref{fig:fit_he_rel_xil_zg_ion1_errors} shows the best fitting model together with the unfolded data (upper panel). The residuals (bottom panel) do not present any significant structure and the fit has  $\chi^2/{\rm d.o.f.} $ of $ 116/124$. 
Here, and throughout this paper, we have fixed the high energy cut-off, as this quantity cannot be constrained in the $<10$ keV energy range of \textit{XMM-Newton}, to $E_{\rm cut}= 300$ keV (in the observer frame), and we have also fixed the inclination angle to $i=30^\circ$ (following for both of these parameters \citealt{Keek2016}). 
We fix the dimensionless black hole spin parameter to $a=0.998$ and allow the disc inner radius to be a free parameter 
(constrained to be larger than the ISCO radius). 

The best fitting model parameters are listed in Table~\ref{tab:parameters}. We obtain a larger disc inner radius than the fit of \cite{Keek2016} to the same data, and a smaller (more plausible) relative iron abundance. Since they used \textsc{relxill} \citep{Dauser2013,Garcia2014}, the two differences between our model and theirs are that we assume a lamppost geometry as opposed to parameterizing the reflection emissivity profile with a broken power-law, and that we self-consistently account for a radial disc ionization profile instead of assuming a single value of the ionisation parameter.

Our fit requires a large value of the `boost' parameter, $1/\mathcal{B}$. This sets the normalization of the relativistic reflection spectrum relative to the direct emission. For $1/\mathcal{B}=1$,
the relative normalization of the reflection component is set entirely by the lamppost geometry and general relativistic light bending, 
whereas $1/\mathcal{B}>1$ returns a stronger reflection component than expected in the lamppost 
geometry and $1/\mathcal{B}<1$ corresponds to a weaker than expected reflection (see \citealt{Ingram2019} for more details). 
A deviation from the expected isotropic lamppost emission can be due to the true source geometry being something other than point-like
and/or an intrinsic velocity of the plasma in the corona
that beams the emission. 
Since the physical assumptions of \textsc{reltrans} are only valid for a boost parameter reasonably close to unity, we impose a hard upper limit in our fits of $1/\mathcal{B}=3$ to avoid unrealistic values. We see that the best fitting value in our fit is close to this upper limit. 
Interestingly alongside with the large boost value the fit requires a low iron abundance. Therefore even though the relativistic reflection component is fundamental to fit to the data, the iron line does not seem prominent enough to require solar iron abundance. A similar behaviour has been found by \citet{Shreeram2020} for a black hole binary.

\begin{table}
\renewcommand{\arraystretch}{1.5}
\caption{Best fitting parameters for the different models in the paper. First column lists the free parameters in the model; a dash in the model column indicates that the parameter is not part of that model. All models include
Galactic absorption with hydrogen column density fixed to
$n_H = 3.6 \times 10^{20} {\rm cm}^{-2}$, a high energy cut-off  fixed to $300$ keV,  and a narrow Gaussian line in the time-averaged spectrum fixed at $7.01$ keV in the galaxy rest frame, with cosmological redshift $z= 0.025785$. The black hole is considered maximally spinning ($a=0.998$) in all relativistic models. Models [A] and [B] fit to the time-averaged spectrum only. Model~[A] does not include the relativistic reflection component, Model~[B] does. Model~[1] fits two lag spectra in the frequency range $0.02-0.7$ mHz. Models~[2] fits the complex cross-spectra in the same frequency ranges simultaneously to the time-averaged spectrum, whereas the last column model fits just the complex cross-spectrum.
Errors are all $90\%$ confidence. }
\begin{tabular}{  p{1.2cm} |m{0.8cm} m{0.8cm} |  m{0.8cm} |m{0.8cm} }
  \hline
  \multicolumn{1}{c} {} & \multicolumn{2}{c}{Time-ave spec} & \multicolumn{1}{c}{\makecell{Time-ave \& \\ Lag spec }}    & \multicolumn{1}{c}{\makecell{ Time-ave \& \\Cross spec }}  \\
  \hline
\multicolumn{1}{c} {Parameter} &
\multicolumn{1}{c} { \makecell{\textsc{pw} + \textsc{xil} \\ $\mathrm{[A]}$}} &
\multicolumn{1}{c} { \makecell{\textsc{rel} + \textsc{xil}\\ $\mathrm{[B]}$}} & 
\multicolumn{1}{c}{\makecell{ \textsc{model} $ [1] $ } }& 
\multicolumn{1}{c}{\makecell{\textsc{model}  $ [2] $}} \\  
  \hline

$h \,[R_g]$               & -                       & $13^{+13}_{-7}$        &  $ 11.4^{+42.2}_{-6.6} $  & $12.4^{+25.6}_{-9.6} $  \\
$ r_{\rm in} [R_g] $     & -                       & $17^{+6}_{-6}$         & $17.6^{+6.1}_{a} $        &  $ 17.4^{+10.9}_{-16.2}  $ \\
$\Gamma_{\mathrm{pl}}$  & $2.19^{+0.02}_{-0.03}$  & $ 3.00^{+0.02}_{-0.08} $ & $2.29^{+0.07}_{-0.08} $ &   $ 2.29^{+0.07}_{-0.08} $   \\
$\log \xi$         & $0$ (\textsc{xil})      & $ 1.7^{+0.3}_{-0.4}$   &  $1.5^{+1.1}_{-1.1} $ &  $1.6^{+1.1}_{-1.6} $   \\
A$_{Fe}$                  & $0.5^{+0.1}_{b}$        & $0.6^{+0.1}_{-0.1} $  & $0.62^{+0.17}_{b}$   &  $ 0.65^{+0.17}_{b}$  \\
$1/\mathcal{B} $ (Boost)           & -                       & $ 2.6^{c}_{-1.4}$    & $ 2.7^{c}_{-1.5} $ &   $2.62^{c}_{-1.3} $   \\
Mass  [$10^6M_{\odot}$]   & -                       & -                     & $ 1.13^{+2.0}_{-0.7}$   &  $0.4^{+1.8}_{d} $  \\
$\chi^2/\mathrm{d.o.f.}$              & $ 168/127 $           & $ 116/124 $            & $128/144 $      &   $ 144/136 $   \\
  \end{tabular}
  \vspace{5pt}
	\begin{list}{}{}
		\item[$^a$] The lower limit of the inner radius is the ISCO ($1.237\,R_{\rm g}$).
 		\item[$^b$] The lower limit of the iron abundance is $0.5$. 
    	\item[$^c$] The upper limit of the boost parameter is $3$
    	\item[$^d$] The lower limit of the mass is $1\,M_{\odot}$
	\end{list}
    \label{tab:parameters}
 \end{table}

\subsection{Time-averaged and lag spectrum}

We now model the lag-energy jointly with the time-averaged spectrum, again using \textsc{reltrans}\footnote{The configuration of the radial zones in the model is the same as that used for the time-averaged spectral fitting.}.  
We again ignore energies $< 2$ keV in the time-averaged spectrum because of the complicated structure of the warm absorber. For the lags, we instead consider the extended energy range $0.3 - 10.0 $ keV. 
Although soft lags can be caused by the response of the warm absorber to changes in the ionizing flux (\citealt{Silva2016}) which is not accounted for in our model, these lags should only be relevant for variability on longer timescales than the Fourier frequency ranges considered in our analysis. We therefore choose to model the reverberation lags down to $0.3$ keV in order to maximize the signal to noise of the data. 

The first model, hereafter Model [1], considers the time-averaged spectrum and lags calculated in the two Fourier frequency ranges where the source shows high coherence. 
Since the higher frequency range ($0.2 - 0.7$ mHz) seems not to be affected by the hard lags (\citealt{DeMarco2013a,Kara2013c}) 
we use the pivoting prescription only in the lower frequency range ($0.02-0.2$ Hz). 
Thus, in the higher frequency range only the normalisation of the illuminating power law is varying.
We note that when pivoting {\it is} included in this frequency range, the hard lags also fit to the reverberation lags, biasing our results. 
This happens because even though the iron line signal in the lag energy
spectrum is significantly higher than the underlying continuum signal, 
it is not sufficiently strong to dominate the fit. 

\begin{figure}
	\includegraphics[width=\columnwidth]{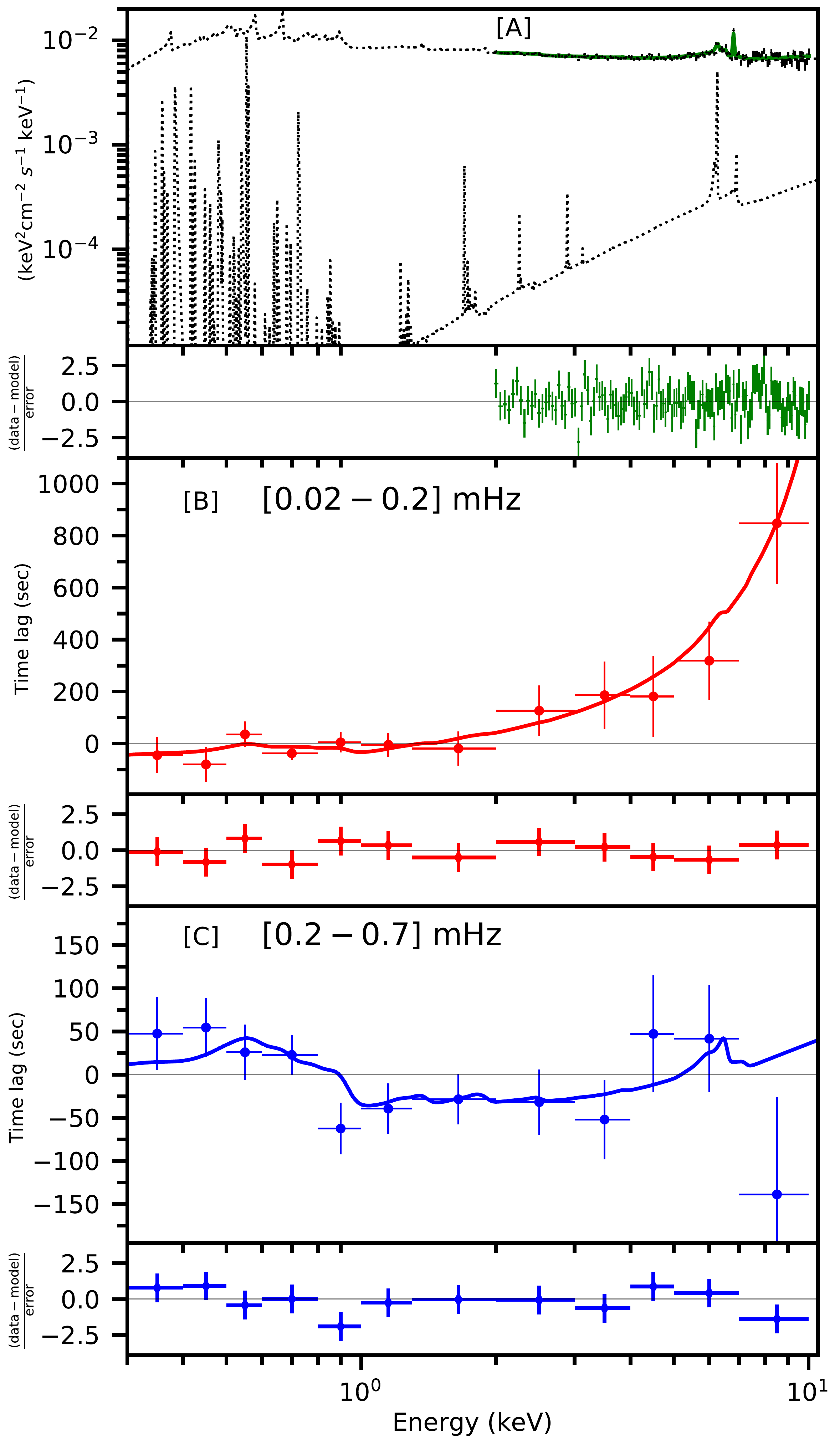}
    \caption{Joint fit of the time-averaged energy spectrum and the lag energy spectra in different Fourier frequency ranges (specified in the panels).  Panel~A: unfolded time-averaged spectrum, the dashed curves are the model components; panels B and C: time lag spectra. Below every panel we show the residuals of the fit. The time-averaged spectrum is fitted in the $2-10$ keV energy range, the time lag spectra in $0.3-10$ keV. The model of the first frequency range accounts for both illuminating continuum and reverberation lags whereas only the reverberation lags are accounted in the second frequency range. 
    Model curves are plotted with higher resolution than the data for clarity.}
    \label{fig:DC_lag01}
\end{figure}

The best fitting parameters of  Model [1]
are listed in Table~\ref{tab:parameters}, and the best fitting time-averaged energy spectrum and lag spectra are shown in panels A, B, C in Fig~\ref{fig:DC_lag01} alongside the data.
The dashed lines in panel A are the different components of the time-averaged spectrum. 
The fit has an acceptable $\chi^2/{\rm d.o.f.}$ of $128/144$.
The model has been plotted with higher resolution than
the data for clarity, which allows the small reverberation feature to be seen in the low frequency range of the model 
(see also \citealt{Mastroserio2018}),
which is unfortunately impossible to detect with the available data resolution. 

\subsection{Time-averaged and complex cross-spectrum}
Finally we attempt to account for the variability amplitude of the reflection component in addition to the time lags by simultaneously fitting the complex cross-spectrum as a function of energy jointly with the time-averaged spectrum. We now only fit in the $2-10$ keV energy range, since the warm absorber will affect the energy dependence of the cross-spectrum even if it does not contribute to the time lags. 

\begin{figure*}
	\includegraphics[width=\textwidth]{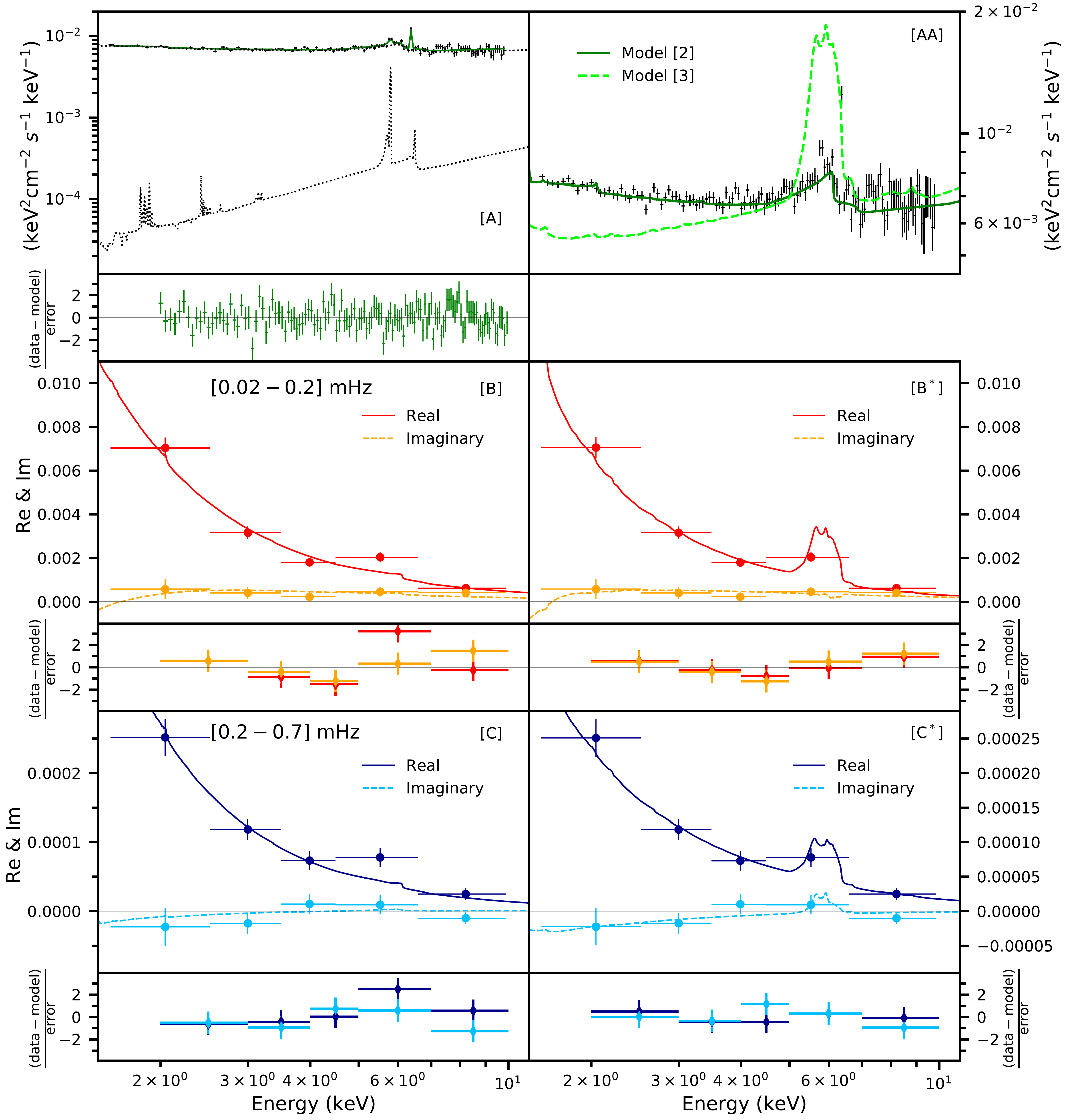}
    \caption{Left panels: unfolded spectra of the time-averaged energy spectrum and the real and imaginary part of the cross-spectrum with the best fitting Model~[2]. Panel~A shows the time-averaged spectrum (solid line) and the different components (dotted lines). Panels B to C show the complex cross-spectrum  in 2 ranges of frequencies (specified in the panels). Below every panel there are the residuals of the best fitting model. 
    Right panels: Panels B$^*$ to C$^*$ show the unfolded real and imaginary part of the cross-spectrum with the best fitting Model~[3] that does \textit{not} consider the time-averages spectrum (the frequency ranges are the same of the left-side panels). In all the cross-spectrum panels solid lines model the real part and dashed lines model the imaginary part.
    The top right panel AA shows the unfolded time-averaged spectrum with Model~[2] (solid dark green line) superimposed to the time-averaged spectrum calculated from the best fitting parameters of Model~[3] (dashed light green line).
    For all the panels the pivoting prescription has been accounted for only in the lowest frequency range. 
 }
    \label{fig:DC_freq01}
\end{figure*}

We fit simultaneously to the time-averaged spectrum and the complex cross-spectrum in the same two frequency ranges. This model, hereafter Model~[2], is therefore the same as Model~[1], except the correlated variability amplitude is considered in addition to the time lags and the time-averaged spectrum. 
The best fitting model is shown in the left panels A, B and C of Fig.~\ref{fig:DC_freq01}, where the upper panel shows the unfolded time-averaged spectrum and the bottom panels the unfolded real (solid lines) and imaginary (dashed lines) parts of the cross-spectrum.  
Again the hard lag produced by the pivoting power-law is enabled only for 
the lower frequency range. 
We note that the iron line signal, represented by a single data point, is too weak in the model in both the complex cross-spectra. 
Even though the $\chi^2/{\rm d.o.f.}$ of $ 144/136 $ is still formally acceptable, the
structured residuals in the iron line range indicate our model is not able to properly fit  the time-averaged spectrum and complex cross-spectra simultaneously. 
It is worth noting that the parameters of the fit are similar to those of the previous fit (Model [1]), apart from the black hole mass which is poorly constrained in this case with
only an upper limit of $\sim 2.2 \times 10^{6} M_{\odot}$. 
This is because the model does not reproduce the reverberation 
feature in the lag energy spectrum that provides the mass constraint. 

\subsection{Cross-spectrum model}

As an experiment to determine what is driving our Model~[2] fit, we fit the cross-spectrum in both frequency ranges without considering the time-averaged spectrum. 
This is instructive because the time-averaged spectrum tends to dominate over the cross-spectrum in the fit due to its far better statistics. We refer to this experiment as Model~[3].
Right panels B$^*$ and C$^*$ of Fig.~\ref{fig:DC_freq01} show the resulting best fitting model, which has a reduced $\chi^2$ of $ 8.78/8 $ and exhibits a prominent iron line feature in both frequency ranges. 
In panel AA of Fig.~\ref{fig:DC_freq01}, instead, we plot the time-averaged spectrum corresponding to best fitting parameters of Model~[3] (light green dashed line). This line has not been fitted to the data but only superimposed to the time-averaged spectrum.
In the same panel as a comparison we plot the Model~[2] time-averaged spectrum (dark green solid line). 
We see that the two spectra are dramatically different, with the cross-spectrum-only fit predicting a much stronger and narrower iron line (resulting from a much larger source height $\sim 44 R_{\rm g}$ and iron abundance $\sim 10$). 
We therefore see that the model can either reproduce the observed cross-spectrum or the observed time-averaged spectrum. The far better statistics of the time-averaged spectrum drive the fit to favour it, and therefore miss the iron line features in the cross-spectrum.
This discrepancy is very similar to what \citet{Zoghbi2020} found in NGC~5560, even though  
in that case the problem already appeared fitting the lag spectrum.
They found that using a static lamppost geometry model over-predicts the iron line 
feature in the time-averaged spectrum. 
Although our model is able to jointly fit lag and time-averaged spectra, when we consider the correlated 
variability amplitudes we encounter the same problem (compare their Fig.7 with our panel AA of Fig.~\ref{fig:DC_freq01}). 
In order to fit the cross-spectrum our model maximises the iron abundance in the accretion disc which leads to an unrealistic iron line feature in the time-averaged spectrum. 
In our case it seems that the lags are compatible with the parameters of the time-averaged spectrum, however, the data requires more variability in the iron line compared to what is predicted by the model.
\citet{Zoghbi2020} argue the assumed static lamppost geometry  could be the cause of this discrepancy. 
We expand on this argument in Section~\ref{sec:discussion} also considering our results on the black hole mass. 

\subsection{Model parameters}
\subsubsection*{Inner radius and ionisation profile}
When we consider only the  time-averaged spectrum (Model~[B]) the inner radius is $17\pm 6$ $R_{\rm g}$.  
This value is larger than that found by \cite{Keek2016}, who fixed $r_{\rm in}$ to the ISCO and inferred a black hole spin of $a=0.89\pm 0.05$ from the time-averaged spectrum of the same observation that we consider here. 
They used \textsc{relxill} for their fit, as noted in Section~\ref{sec:DC}. The configuration of \textsc{reltrans} we use here is different in two ways:
it uses an emissivity profile calculated in the lamppost geometry instead of parameterizing it with a broken power-law function, and it considers a radial ionization profile instead of a constant ionization parameter. 
\citet{Shreeram2020} showed that assuming a radial ionisation profile in the accretion disc improves the spectral fit compared to a constant ionisation and 
the value of the inner disc radius depends on the nature of this profile. 
The radial ionisation profile quite dramatically changes the shape of the iron line, in particular the red wing (see Fig.~9 in \citealt{Ingram2019}), requiring less broadening from the relativistic effects. 
In order to calculate the radial ionization profile, our model 
assumes the density profile corresponding to zone~A of the \citet{Shakura1973} disc model,
in which the dominant sources of pressure and opacity are, respectively, radiation and electron scattering.  
According to \citet{Shakura1973}, the outer limit of zone~A in an AGN with a $0.5 - 10$ keV luminosity of $~5 \times 10^{43}$ erg/s \citep{Grupe2007}, and assuming viscosity parameter $\alpha = 0.01$ and black hole mass  $10^6 $ or $50\times 10^6$ $M_{\odot}$, is $R_{\mathrm{ab}} \sim 1000~R_g $ or $ \sim~72 R_g $ (increasing $\alpha$ increases $R_{\mathrm{ab}}$). Thus we can comfortably consider most of the emission to come  from a zone~A density profile accretion disc. 
We note that the disc scale height in the zone A regime can be significantly greater than zero (see e.g. \citealt{Taylor2018a}), whereas here we assume an infinitely thin disc. 
Since the disc scale height can impact on the reverberation signal \citep{Taylor2018b}, we plan to include a more general disc geometry in our model in future.
Including the lag spectra in the fit (Model [1]) does not change the best fitting inner radius. 
However, the lower limits are now compatible with ISCO within 90\% confidence, because the broad and strong iron line observed in the lag spectrum favours a smaller inner radius.

\subsubsection*{Iron abundance}

It is also notable that the iron abundance yielded by our models, $A_{\rm Fe}\sim 0.5-0.7$, differs significantly from the large value of $A_{\rm Fe} = 3.9 \pm 0.7$ returned by the fits of \cite{Keek2016}. It is common for X-ray reflection modelling to measure significantly super-solar iron abundances both for AGN and black hole X-ray binaries (e.g.  \citealt{Dauser2012, Parker2015, Garcia2015}; see \citealt{Garcia2018} for a brief review). 
Again it is possible that including a radial ionization profile goes some way to reducing the inferred iron abundance, as speculated by \cite{Ingram2019}.

\subsubsection*{MCMC simulation}

\begin{figure*}
	\includegraphics[width=\textwidth]{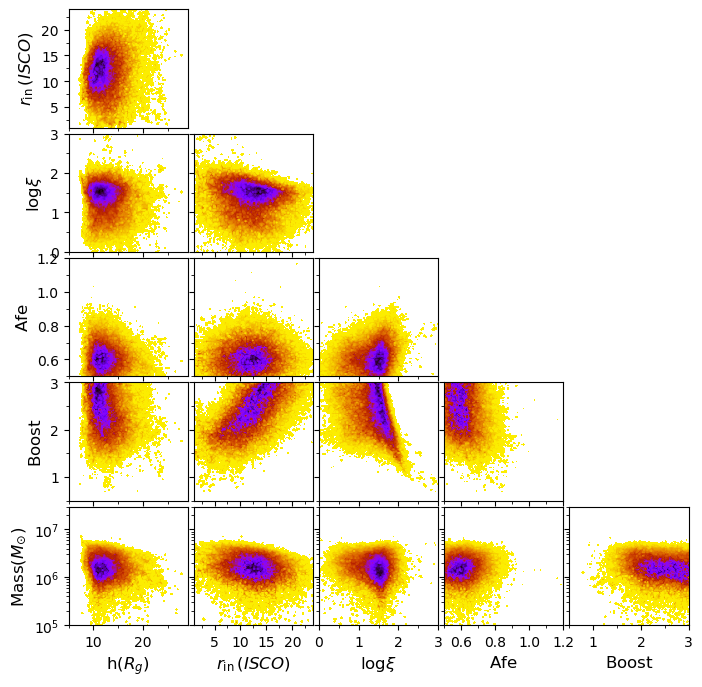}
    \caption{Probability distribution from the MCMC simulation of Model~[1] (see text for the details). The purple, orange and yellow are $1$, $2$, $3\sigma$ contours respectively. The inner radius is expressed in units of ISCO which is $1.237\,R_{\rm g}$ in this case.
 }
    \label{fig:corner_plot}
\end{figure*}

When we include the variability amplitude by fitting the complex cross-spectrum instead of the lag spectrum (Model [2]), although we achieve a statistically acceptable $\chi^2$, we see clear residual structure in the iron line region and therefore do not consider this model to adequately describe the data. 
We therefore focus our further analysis on the results of fitting the time-averaged spectrum together with the lag spectrum (Model~[1]). In order to probe the parameter space of this model and 
look for parameter correlations we run a Markov Chain Monte Carlo (MCMC) simulation in \textsc{xspec} with $200$ walkers and a total length of $1$ million steps.
The walkers start from the Model~[1] best fit and we set a burn-in
period of $5000$ steps to assure the convergence of the chain. 
Fig.~\ref{fig:corner_plot} shows the integrated probability distribution for some of the model parameters, the purple regions represent the $1\sigma$ contour, orange and yellow $2$ and $3\sigma$. 
The inner radius is strongly positively correlated with the boost parameter.
The reason for this is that both contribute to the strength 
of the reflection emission in the opposite sense:
when the inner radius is closer to ISCO the reflection is stronger and a lower value of the boost is required.
However, the best fit value of the boost parameter is significantly above $1$ which means that the data require a stronger reflection component than the one produced by the lamppost model. 
This can be interpreted as the velocity of the plasma in the lamppost being non-zero and  pointing towards the black hole, boosting the flux  illuminating the disc. 
Contrary to what has been seen in previous works the black hole mass is not  degenerate with the height of the lamppost (e.g. \citealt{Cackett2014}) nor with any other parameter 
(\citealt{Mastroserio2019} found a correlation with the ionisation parameter). 
This is likely due to the wide range of timescales and the simultaneous fit to the time-averaged spectrum that we considered in the current work.
In Section~\ref{sec:discussion} we further study the relation between mass and height of the lamppost. 

\section{Discussion}

In this work we have jointly modelled the time-averaged X-ray spectrum and the
short timescale X-ray variability of the Seyfert galaxy Mrk~335 using our relativistic reflection model \textsc{reltrans}. 
We considered the high-flux XMM-Newton observation from 2006, in which 
\citet{Kara2013c} found evidence of reverberation lags in the $\sim 0.2-0.7$ mHz Fourier frequency range. 
We first modelled the time-averaged energy spectrum alone (Model [B]) before additionally modelling the lag-energy spectrum simultaneously in
two frequency ranges ($0.02 - 0.2$ mHz and $0.2 - 0.7$ mHz; Model~[1]), with our model for the lowest frequency range
including a prescription to account for the intrinsic hard lags
(\citealt{Mastroserio2018,Mastroserio2019}).
We then extended our analysis to 
fit simultaneously to the complex cross-spectrum and the time-averaged spectrum (Model [2]).
We achieve statistically acceptable fits for both Model~[1] and Model~[2]. However, the Model [2] fit suffers from structured residuals around the iron line region, and so here we further discuss the details of our more successful Model~[1] fit as well as the possible reasons for the deficiencies in our Model~[2] fit.

\subsection{Black hole mass}
\label{sec:discussion}
\begin{figure}
	\includegraphics[width=\columnwidth]{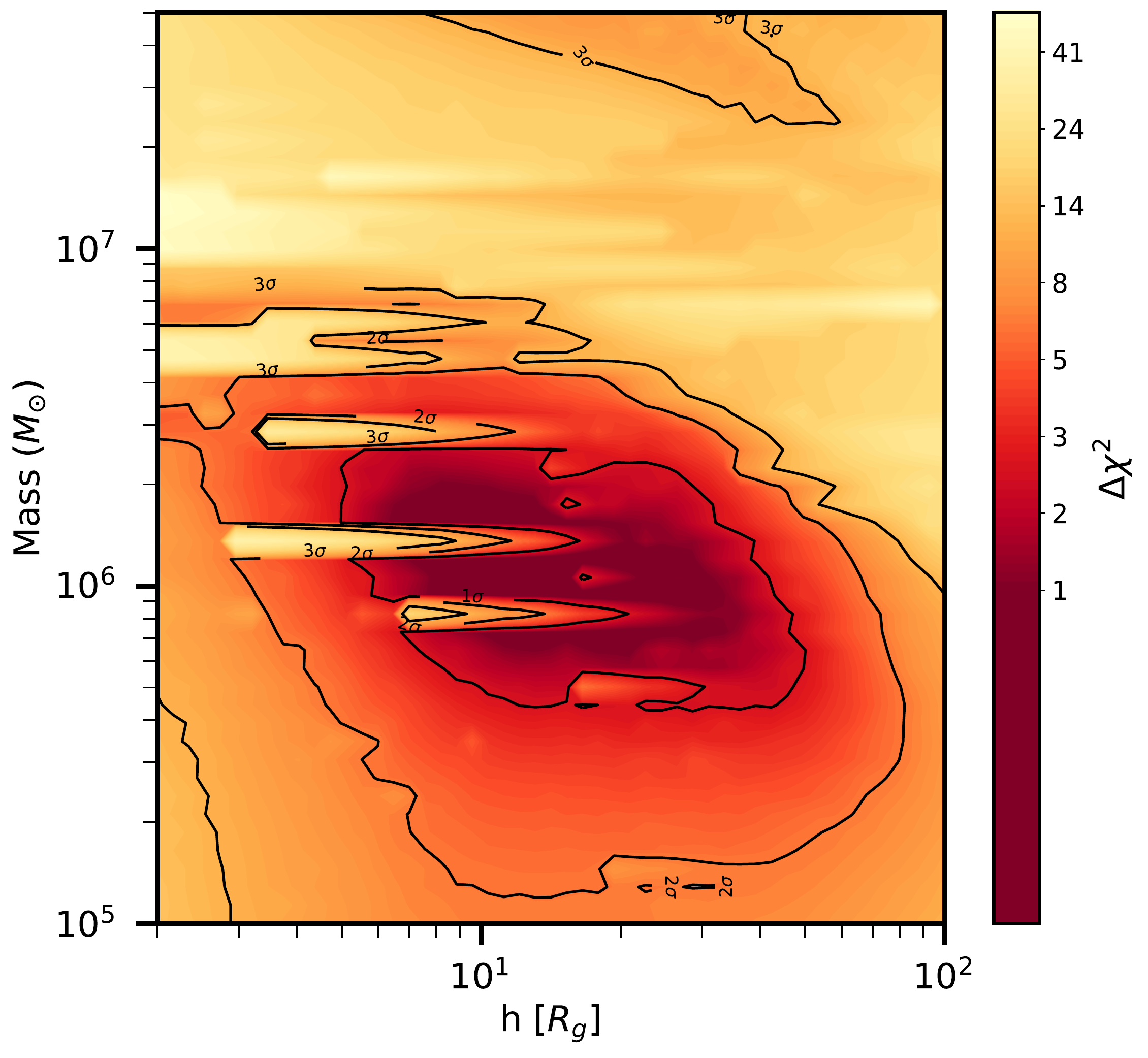}
    \caption{2D $\chi^2$ contour plot of black hole mass and source height resulting from Model~[1].  
 }
    \label{fig:2D_contour}
\end{figure}

The fits that consider timing properties are sensitive to black hole mass. In previous analyses, fits to the lag-energy spectrum have suffered from a degeneracy between the source height and the black hole mass (\citealt{Cackett2014,Ingram2019}). 
This occurs because the same reverberation lag can be reproduced if the source height is many small gravitational radii or a few large gravitational radii. 
We see from the bottom left panel in Fig.~\ref{fig:corner_plot}, 
and from the more detailed contour plot of $\chi^2$ as a function of 
black hole mass and source height presented in
Fig.~\ref{fig:2D_contour}, that there is no such degeneracy in our Model [1] fit. 
This is because we consider two frequency ranges instead of only one. 
This point can be illustrated by comparison with the \textsc{reltrans} fit 
to only the $0.2 - 0.7$ mHz lag-energy spectrum 
of the same observation presented in Fig.~17 of \citet{Ingram2019}. 
In that case, the best fitting black hole mass was $\sim 7 \times 10^6~M_\odot$, but other solutions with mass values as high as $\sim 50 \times 10^6~M_\odot$ were statistically acceptable within $1\sigma$ confidence.
These high mass solutions were in the phase-wrapping regime, 
and are ruled out from our new fit because they predict 
very large reverberation lags in the $0.02 - 0.2$ mHz 
frequency range that are not observed.

\begin{figure}
	\includegraphics[width=\columnwidth,trim=0.0cm 0.0cm 15.0cm 0.0cm,clip=true]{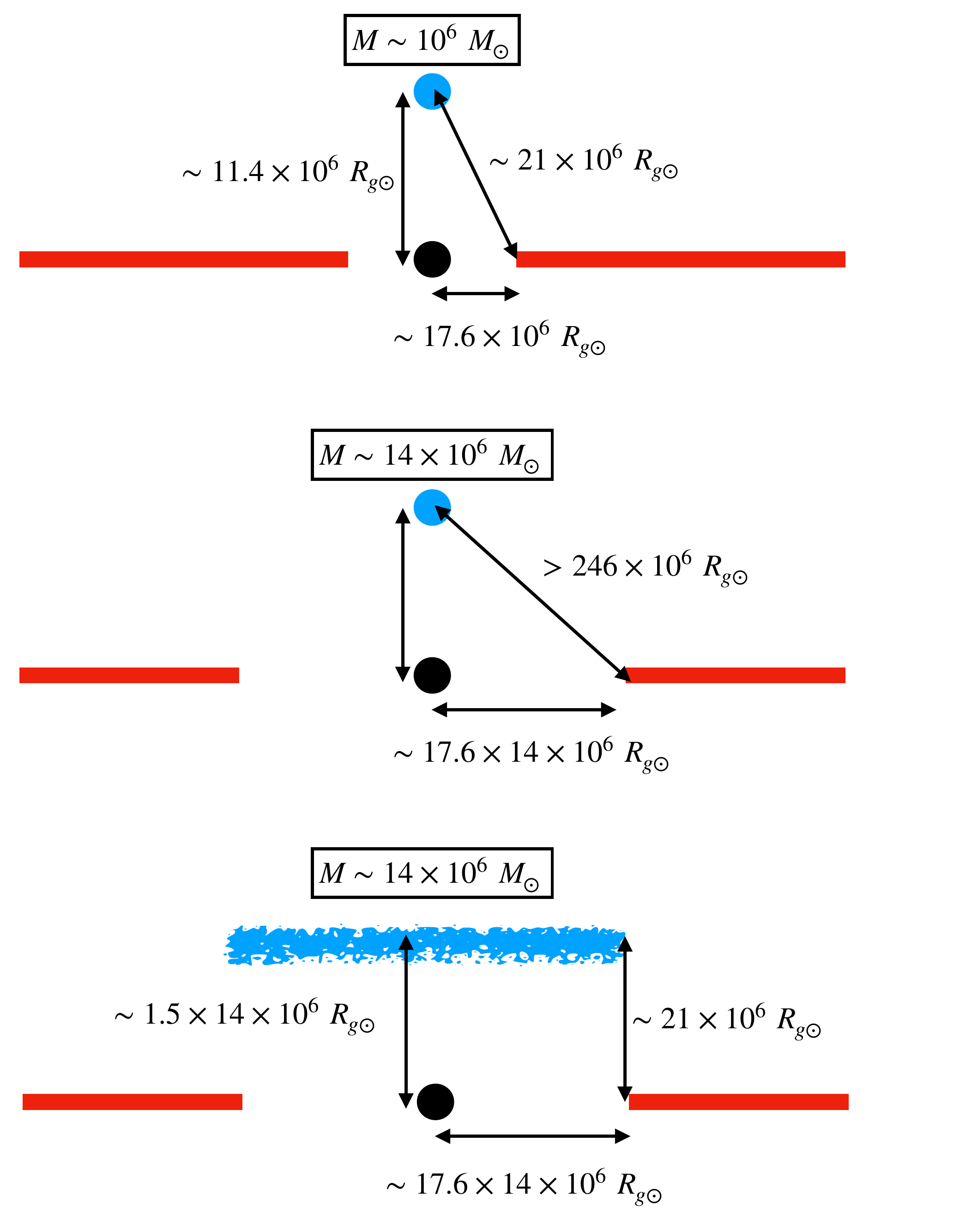}
    \caption{Schematic comparing the lamppost geometry with a radially extended corona geometry ($R_{g\odot}=GM_\odot/c^2$). The top panel represents our Model [1] fit. For the middle panel, all parameters are the same, except the black hole mass is now 14 times larger. The physical distance from the corona to the disc now must be larger than it was for the top panel, no matter how low the source is placed. The bottom panel illustrates a radially extended corona. We see that the same physical distance from the corona to the disc can now be achieved even for a $14\times 10^6~M_\odot$ black hole.}
    \label{fig:schematic}
\end{figure}

Our best-fitting black hole mass of $1.1^{+2.0}_{-0.7} \times 10^6~M_\odot$ (errors are 90\% confidence) 
is roughly one order of magnitude
smaller than the optical reverberation mass measurements ($M=14.2\pm 3.7 \times 10^6M_{\odot}$;
\citet{Peterson2004} and $26\pm 8 \times10^6M_{\odot}$; 
\citealt{Grier2012}). 
These higher mass values are incompatible with our modelling with high statistical confidence since, as discussed above, parameter combinations that can reproduce the $0.2-0.7$ mHz lag-energy spectrum with such high masses 
predict large lags in $0.02 - 0.2$ mHz that are not observed.
This discrepancy is not due to dilution of the reverberation lag by the presence of continuum emission variability with no time lags between energy bands (see e.g \citealt{Uttley2014}) because dilution is fully accounted for in our model -- although there could be an extra source of dilution that is not in our model. 
The discrepancy could be because we assume a lamppost corona, if in reality the corona is extended (as was suggested by \citealt{Wilkins2015} for this very observation). 
The schematic in Fig.~\ref{fig:schematic} illustrates how an extended corona model could in principle accommodate a higher mass than the lamppost model. 
The top panel illustrates our best fit, whereby the reverberation lag roughly results from the $\sim 21~R_g$ distance from the corona to the disc. The second panel shows that the same physical distance from corona to disc cannot be reproduced if the mass is increased by a factor of 14. 
This is because the physical size of the $\sim 17.6~R_g$ disc inner radius is now a factor of 14 larger (and the inner radius in units of $R_g$ is constrained by the iron line profile). The third panel illustrates that a distance of $\sim 21 \times 10^6~R_g$ \textit{is} possible for an $M\sim 14 \times 10^6~M_\odot$ black hole with an extended corona at a height of $\sim 1.5~R_g$.

Previous X-ray reverberation analyses of Mrk~335 have also inferred a higher black hole mass than what we find here. 
\cite{Chainakun2016a} fit to the lag-energy spectrum of the same observation that we consider here. Their model assumes a lamppost geometry and is largely similar to ours, with some small differences such as the form of the radial ionization profile; their best fitting mass value was $M \sim 13.5 \times 10^6~M_\odot$. 
The major differences between their analysis and ours are that they only considered one frequency range, whereas we consider two (their one frequency range corresponds to our higher frequency range),
and that we conduct a more thorough exploration of parameter space. 
Given that the single frequency fit presented in \citet{Ingram2019} included statistically acceptable phase-wrapped solutions with high black hole mass, 
it could be that the \cite{Chainakun2016a} fit is also in the phase-wrapping regime. 
If this is the case, then the high black hole mass returned by their fit would also be ruled out upon consideration of a second, lower frequency range. 
\citet{Emmanoulopoulos2014} used a different procedure of fitting the lag-frequency spectrum between the $0.3-1$ keV and $1.5-4$ keV energy bands and obtained a mass of $M \sim 20 \times 10^6~M_\odot$. However, they did not simultaneously fit to the time-averaged spectrum, and they simply assumed that respectively $30\%$ and $0\%$ of the flux in the $0.3-1$ keV and $1.5-4$ keV bands is due to reflection, whereas we include a full model of the rest frame reflection spectrum. 

It is also worth investigating whether the assumptions going into the optical reverberation mass measurements can be reasonably changed
in order to obtain values compatible with our X-ray reverberation measurement.
Optical reverberation mapping consists of measuring the time delay $\tau$ 
between variations in the optical continuum, 
assumed to come from the accretion disc, 
and corresponding variations in the broad optical emission lines, 
assumed to come from clouds orbiting the black hole at a distance 
far larger than the accretion disc in the so-called broad line region (BLR). 
Assuming that $\tau$ is a light crossing delay, and assuming that the measured (full width at half maximum)
velocity width of the lines, $V_{\rm FWHM}$, results from virialized rotation 
around the black hole, the black hole mass is given by
\begin{equation}
    M = f \frac{ c \tau (V_{\rm FWHM}/2)^2}{G}.
\end{equation}
Here, $f$ is a factor that depends on
the structure, kinematics and 
orientation of the BLR ($f=3$ for a BLR consisting of a thin spherical shell with an isotropic velocity distribution; \citealt{Onken2004}). 
\citet{Peterson2004} set $f=5.5$ for their mass measurement of 
$M=14.2\pm 3.7 \times 10^6M_{\odot}$ for Mrk 335, 
which is the value that results from the reasonable assumption that AGN and quiescent galaxies follow on average the same
relationship between black hole mass and bulge stellar velocity 
dispersion \citep{Onken2004}. It is in principle possible to adjust the assumptions of 
optical reverberation mapping to obtain $M\sim 10^6~M_\odot$ for Mrk 335, but 
the required adjustment,  setting $f=0.4$, therefore seems unlikely.
Moreover, such a low mass is in 
tension with the measured $5100$~Å flux, from which \citet{Peterson2004} 
infer a bolometric luminosity of $\sim 6.5 \times 10^{44}~{\rm erg}\, {\rm s}^{-1}$. 
This luminosity is equal to the Eddington limit for a black hole of mass $M\sim 4 \times 10^6~M_\odot$, which roughly coincides with our $90\%$ confidence upper limit. 
Thus our mass value implies Eddington limited or mildly super-Eddington accretion in Mrk~335, 
which is not impossible but has not previously been thought to be the case for Mrk~335.

\subsection{Coronal Geometry}

Although we have achieved a good fit for the time lags, we are skeptical of our fit for the full cross-spectrum because it suffers from structured residuals around the iron line region (although formally the fit is still statistically acceptable). It therefore seems that the model has problems reproducing the energy-dependent correlated variability amplitude. This could again potentially be because we assume a lamppost geometry, if in reality the corona is extended. \citet{Wilkins2015} inferred a radial extent of $\sim 26~R_g$ during the high-flux epoch that we consider, followed by a contraction when the flux decreased. 
In such an extended corona, intrinsic hard lags may arise due to inward propagation of accretion rate fluctuations \citep{Wilkins2016}, which we instead account for with a fluctuating power-law index. 
Our prescription provides a good mathematical description of fluctuations propagating through a disc and into a compact corona (Uttley \& Malzac in prep.), 
but likely differs significantly from propagation in an extended corona. The reverberation signal itself would also differ significantly from the lamppost case (see e.g. Fig. \ref{fig:schematic} but also \citealt{Wilkins2016}).

We note that our analysis is the first ever attempt to jointly fit the full energy-dependent cross-spectrum and time-averaged spectrum of an AGN. However, we have previously applied the same analysis to the black hole X-ray binary Cygnus X-1 \citep{Mastroserio2019}, and in that case the model worked very well. We cannot currently rule out this being due to the two systems having different coronal geometries, with Cygnus X-1 having a more compact corona than Mrk 335 during the observations we analysed. 
Alternatively, the two sources could have roughly similar extended coronal geometries, with our lamppost model better able to approximate the true signal for the lower frequencies (in terms of $R_g/c$) typically probed by X-ray binaries than the higher frequencies (in terms of $R_g/c$) probed by AGN. High signal to noise \textit{NICER} observations of X-ray binaries, for which timing properties can be constrained up to frequencies as high as $\sim 30$ Hz, may enable this hypothesis to be tested. The signal for AGN may also be more significantly modified by absorption and winds, given the very different environment associated with the two classes of object and the different opacity in the disc (e.g. \citealt{Miller2010, Mizumoto2018})

\section{Conclusions}

The results of our X-ray spectral-timing analysis of Mrk~335 using the model \textsc{reltrans} can be summarised as follows: 
\begin{enumerate}
    \item Our fits return a best fitting disc inner radius that is truncated outside of the ISCO for any value of black hole spin ($R_{\rm in} \approx 17~R_g$), and a sub-solar iron abundance ($A_{\rm Fe}<0.7$ ). Although our fits that include timing properties are consistent with $R_{\rm in}$ at the ISCO within $90\%$ confidence, our results differ from previous spectral fits for the same observation \citep{Keek2016}, which returned a small disc inner radius and a super-solar iron abundance. 
    We attribute this to the more realistic treatment of the ionisation structure of the accretion disc in our model, allowing for a radial profile rather than assuming a constant ionisation parameter. This suggests that models with a radial ionization profile may return larger disc inner radii and smaller iron abundances for other AGN than the constant ionisation models that have mainly been used in the past. 
    \item Our simultaneous fit to the time-averaged spectrum and the lag-energy spectrum (Model~[1]) returns a black hole mass of $M=1.1^{+2.0}_{-0.7} \times 10^6~M_\odot$, which is not compatible with UV/optical reverberation mapping results ($14-26 \times 10^6~M_\odot$). We argue that our use of the lamppost geometry might be responsible for this low mass value, whereas an extended corona model may return a larger mass.  Proper modelling of an extended corona is necessary to confirm this claim.
    \item Our model is not satisfactorily able to simultaneously describe the energy dependent correlated variability amplitude of Mrk~335 and the time-averaged spectrum. Fitting to the cross-spectrum alone returns parameters that predict a very strong iron line in the time-averaged spectrum that is not observed. Again this seems to challenge the lamppost geometry, as it has been already suggested by \citet{Wilkins2015}. 
    An extended corona could provide more correlated variability amplitude in the reflection response due to the different illumination of the accretion disc, without significantly changing the lags because the distance between the corona and the disc is similar to the lamppost case. 
\end{enumerate}

\section*{Acknowledgements}
The authors would like to acknowledge the anonymous referee for the very helpful comments and suggestions. G.M. acknowledges the support from NASA grant 80NSSC19K1020 and together with M.K. the support from NWO Spinoza. A.I. acknowledges the Royal Society.

\section*{DATA AVAILABILITY}
The data underlying this article are available from the XMM-Newton science archive (\url{http://nxsa.esac.esa.int/}).



\bibliographystyle{mnras}
\bibliography{library_2020} 



\appendix

\section{Transfer functions}
\label{sec:transfer}

The three transfer functions are
\begin{eqnarray}
W_0(E,\nu) &=& \int_{\alpha,\beta} g_{\rm doz}^3 \epsilon {\rm e}^{i 2\pi (1+z)\tau \nu} \mathcal{R}(E_d) d\alpha d\beta \\
W_1(E,\nu) &=& \int_{\alpha,\beta} g_{\rm doz}^3 \epsilon {\rm e}^{i 2\pi (1+z)\tau \nu} \ln g_{\rm sd} \mathcal{R}(E_d) d\alpha d\beta \\
W_2(E,\nu) &=& \int_{\alpha,\beta} g_{\rm doz}^3 \epsilon {\rm e}^{i 2\pi (1+z) \tau \nu} \frac{\partial \mathcal{R}}{\partial \Gamma}(E_d) d\alpha d\beta.
\end{eqnarray}
Here, $\mathcal{R}(E)$ is the restframe specific intensity of reflected emission emergent from the disc and $g_{doz}=g_{do}/(1+z)$, where $g_{do}$ and $g_{sd}$ are respectively the blueshift experienced by a photon travelling from the disc to infinity and from the point source to the disc in an asymptotically flat and stationary spacetime. $\alpha$ and $\beta$ are the horizontal and vertical impact parameters at infinity, such that a given patch of the disc appears on the image plane to be centered on coordinates $(\alpha,\beta)$ and subtends a solid angle $d\alpha d\beta / D^2$ where $D$ is the distance between the observer and the black hole. $\tau$ is the time delay between the arrival at the observer plane of photons direct from the corona and those that reflected off a disc patch specified by the coordinates $(\alpha,\beta)$. $\epsilon$ is the emissivity, which is a function of radius only (see Equation 20 of \citealt{Ingram2019}). Note that these expressions explicitly include cosmological time dilation due to the factor of $(1+z)$ in the index of the exponential.


\bsp	
\label{lastpage}
\end{document}